\pdfoutput=1

\documentclass[12pt,nofootinbib]{article}
\usepackage{amsmath}
\usepackage{amssymb}
\usepackage{graphicx}
\usepackage[colorlinks=true,linkcolor=blue,citecolor=blue]{hyperref}

\usepackage{color}
\input{colordvi.tex}

\numberwithin{equation}{section}

\newcommand{\beq}{\begin{equation}}   
\newcommand{\eeq}{\end{equation}}
\newcommand{\bea}{\begin{eqnarray}}   
\newcommand{\eea}{\end{eqnarray}}
\newcommand{\bear}{\begin{array}}  
\newcommand {\eear}{\end{array}}
\newcommand{\bef}{\begin{figure}}  
\newcommand {\eef}{\end{figure}}
\newcommand{\bec}{\begin{center}}  
\newcommand {\eec}{\end{center}}

\def\GEV#1{10^{#1}{\rm\,GeV}}

\def\lrf#1#2{ \left(\frac{#1}{#2}\right)}
\def\lrfp#1#2#3{ \left(\frac{#1}{#2} \right)^{#3}}
 
\begin{document}

\begin{titlepage}

\setcounter{page}{1} \baselineskip=15.5pt \thispagestyle{empty}

\begin{flushright}
{\footnotesize RESCEU-33/11}\\
{\footnotesize TU-892}\\
{\footnotesize IPMU11-0179}
\end{flushright}
\vfil

\bigskip\
\begin{center}
{\LARGE \textbf{Dark Radiation from  Modulated Reheating}}
\vskip 15pt
\end{center}

\vspace{0.5cm}
\begin{center}
{\large 
Takeshi Kobayashi,$^{a, b, c}$\footnote{takeshi-kobayashi@resceu.s.u-tokyo.ac.jp}
Fuminobu Takahashi,$^{d, e}$\footnote{fumi@tuhep.phys.tohoku.ac.jp}
Tomo Takahashi,$^{f}$\footnote{tomot@cc.saga-u.ac.jp}\\
and Masahide Yamaguchi$^{g}$\footnote{gucci@phys.titech.ac.jp}
}
\end{center}

\vspace{0.3cm}

\begin{center}
\textit{$^{a}$ Canadian Institute for Theoretical Astrophysics,
 University of Toronto, \\ 60 St. George Street, Toronto, Ontario M5S
 3H8, Canada}\\ 

\vskip 4pt
\textit{$^{b}$ Perimeter Institute for Theoretical Physics, \\ 
 31 Caroline Street North, Waterloo, Ontario N2L 2Y5, Canada}\\ 

\vskip 4pt
\textit{$^{c}$ Research Center for the Early Universe, School of
 Science, The University of Tokyo, \\ 7-3-1 Hongo, Bunkyo-ku, Tokyo
 113-0033, Japan}\\ 
 
\vskip 4pt
\textit{$^{d}$ Department of Physics, Tohoku University, Sendai 980-8578, Japan}

\vskip 4pt
\textit{$^{e}$ Institute for the Physics and Mathematics of the
 Universe, The University of Tokyo, \\ 5-1-5 Kashiwanoha,
 Kashiwa, Chiba 277-8582, Japan}\\

\vskip 4pt
\textit{$^{f}$ Department of Physics, Saga University, Saga 840-8502, Japan}

\vskip 4pt
\textit{$^{g}$ Department of Physics, Tokyo Institute of Technology, Tokyo
152-8551, Japan}

\end{center} \vfil

\vspace{0.8cm}

\noindent 

We show that the modulated reheating mechanism can naturally
account for dark radiation, whose existence is hinted by recent observations of
the cosmic microwave background radiation and the primordial Helium
abundance. In this mechanism, the inflaton decay rate depends on a light
modulus which acquires almost scale-invariant quantum fluctuations during
inflation. We find that the light modulus is generically  produced by the
inflaton decay and therefore a prime candidate for the dark radiation. 
Interestingly, an almost scale-invariant power spectrum predicted in the modulated
reheating mechanism gives a better fit to the observation in the presence
of the extra radiation.
We discuss the production mechanism of  the light modulus in detail taking account of
its associated isocurvature fluctuations. We also consider a case where
the modulus becomes the dominant component of dark matter.

\vfil

\end{titlepage}

\newpage
\tableofcontents

\setcounter{footnote}{0}

\section{Introduction}
\label{sec:intro}

There have been accumulating evidences for the existence of extra
radiation.  Recent combined analysis with the data from cosmic
  microwave background (CMB), large scale structure (LSS) and so on
  has given a constraint on the effective number of light neutrino
  species as
  $N_{\rm eff} = 3.86 \pm 0.42$ (1$\sigma$ C.L.)
  \cite{Keisler:2011aw}\footnote{
    Recent another analysis gives $N_{\rm eff} = 4.08^{+0.71}_
      {-0.68}$ at 95\% C.L.  \cite{Archidiacono:2011gq}.
  }, which suggests the presence of extra radiation at about $2\sigma$
  level \cite{Keisler:2011aw,Archidiacono:2011gq,Dunkley:2010ge}. 
Also, it is known that the $^4$He mass fraction $Y_p$ is sensitive to
the expansion rate of the Universe during the BBN epoch~\footnote{
  $Y_p$ is also sensitive to large lepton asymmetry, especially of the
  electron type, if
  any~\cite{Enqvist:1990ad,Foot:1995qk,Shi:1996ic,MarchRussell:1999ig,Kawasaki:2002hq,Yamaguchi:2002vw,Ichikawa:2004pb}.
}, and the authors of Ref.~\cite{Izotov:2010ca} claimed an excess of
$Y_p$ at the $2 \sigma$ level, $Y_p = 0.2565 \pm 0.0010\, ({\rm stat})
\pm 0.0050\, ({\rm syst})$, which can be understood in terms of the
effective number of neutrinos, $N_{\rm eff} = 3.68^{+0.80}_{-0.70}$
$(2 \sigma)$\footnote{
  The authors of Ref.~\cite{Aver:2010wq} estimated the primordial
  Helium abundance with an unrestricted Monte Carlo taking account of
  all systematic corrections and obtained $Y_p = 0.2561 \pm
  0.0108$\,$(68\%{\rm CL})$, which is in broad agreement with the WMAP
  result.
}. Although the precise determination of the Helium abundance is still
limited by systematic uncertainties, it is intriguing that the CMB and
LSS data as well as the Helium abundance from BBN are showing hints for
the presence of additional
relativistic species, $\Delta N_{\rm eff} \sim 1$, while they are
sensitive to the expansion rate of the Universe at vastly different
times.

The extra radiation may be dark radiation composed of unknown particles
(see e.g. Refs.~\cite{Ichikawa:2007jv,Jaeckel:2008fi, Krauss:2010xg,Nakayama:2010vs,deHolanda:2010am,Hamann:2011ge} for particle physics models).  Such particles must be light and have only very weak
interactions with the standard model particles.  Although one can just
add such a light degree of freedom by hand to explain the data, it
still remains a puzzle why there is such a light particle at all, and
how it is produced.  In this paper, we show that there is a well
motivated candidate in a scenario with the modulated reheating
mechanism~\cite{Dvali:2003em,Kofman:2003nx}.

The origin of the density perturbation is one of the unresolved issues
in modern cosmology.  While the inflaton is the prime candidate, there
are several alternatives proposed so far.  The modulated reheating
scenario is one of them, and has been extensively discussed in various
contexts, especially in connection with
non-Gaussianity~\cite{Zaldarriaga:2003my,Suyama:2007bg,Ichikawa:2008ne}.
The key feature of the modulated reheating scenario is that the decay
rate of the inflaton depends on a light scalar field, $\sigma$, which
acquires quantum fluctuations extending beyond the horizon scale
during inflation.
In this paper we focus on the cosmological fate of the scalar
$\sigma$, which was not studied in detail so far.  We find that some
amount of $\sigma$ is necessarily produced by the inflaton decay,
since $\sigma$ must have couplings with the inflaton to make the decay
rate fluctuate. Also, the $\sigma$ has only very suppressed couplings
with the standard model particles since otherwise it would acquire a
thermal mass spoiling the modulated reheating mechanism. 
We therefore argue that the modulus $\sigma$ can naturally account for the extra
radiation. Interestingly, in the presence of the extra radiation, 
observational data favors a larger scalar spectral index $n_s$, and so, 
the modulated reheating mechanism which generically predicts a very flat
power spectrum\footnote{Even though the modulus dynamics itself hardly
contributes to the spectral tilt, we should also note that special
classes of inflationary 
models (such as large-field ones) giving large time-variation to the
Hubble parameter during inflation as large as $|\dot{H}/H^2| \sim
0.01$, can source large tilt for the resulting perturbation power
spectra.\label{foot123}} gives a better fit to observations.
In addition,  a (classical) oscillating background of the modulus can
also play a role of dark matter. We evaluate their abundances and give
some explicit values for an example model. In particular, we show that
the amount of dark radiation can be predicted as $\Delta N_{\rm eff}
\sim 1$ in the model.  

In fact, in this kind of model, (correlated-type) isocurvature
  fluctuations could be generated in dark radiation and/or dark matter
  sectors. Since too large isocurvature fluctuations are inconsistent
  with observations such as CMB, we discuss the conditions to evade
  cosmological constraints for the isocurvature modes.  

\section{Modulated Reheating Mechanism}
\label{sec:2}

The modulated reheating mechanism proposed
in~\cite{Dvali:2003em,Kofman:2003nx} generates cosmological
perturbations through a modulated decay of the inflaton. This can
happen when the couplings between the inflaton and the Standard Model
(SM) particles are determined by a vacuum expectation value of a
modulus field, such as
\begin{equation}
 \mathcal{L} \supset \phi f(\sigma) \mathcal{O}_{\mathrm{SM}}, 
 \label{coupling}
\end{equation}
where $\phi$ is the inflaton, $\sigma$ is the modulus, and
$\mathcal{O}_{\mathrm{SM}}$ represents a SM gauge invariant
operator.\footnote{In~\cite{Chen:2011sja}, a 
light modulus which is coupled with the mass term of the inflaton in a
specific inflation model was considered. However in this case,
even when such a modulus can play a role of dark radiation to some extent,
it is not possible to satisfy the
isocurvature constraint on dark radiation.} 
Given that the modulus is light during inflation, the field
acquires (almost) scale-invariant quantum fluctuations at length
scales that are stretched to super-horizon sizes by the end of
inflation, as
\begin{equation}
 \delta \sigma \;\simeq\; \frac{H_{\mathrm{inf}}}{2 \pi } \label{deltasigma},
\end{equation}
where $H_{\rm inf}$ denotes the Hubble parameter during inflation.
This results in the inflaton possessing different decay rates among
different patches of the universe, producing cosmological
perturbations.
As for perturbations generated by the inflaton itself, we
ignore them (or consider them to be negligibly tiny) throughout this paper.

Let us begin by computing the curvature perturbations~$\zeta$ in the
modulated reheating mechanism. Using the $\delta
\mathcal{N}$-formalism~\cite{Starobinsky:1986fxa,Sasaki:1995aw,Wands:2000dp,Lyth:2004gb},
$\zeta$ can be computed as the difference in the e-folding
number~$\mathcal{N}$ on a constant energy density hypersurface, among
different patches of the universe with different inflaton decay
rates~$\Gamma$.  We consider an inflaton $\phi$ which undergoes 
harmonic oscillations after inflation, so that its energy density
redshifts as that of matter, i.e. $\rho_\phi \propto a^{-3}$ (here $a$
is the scale factor of the universe). Assuming a sudden decay of the
inflaton into radiation ($\rho_r \propto a^{-4}$), then the number of
e-folds obtained between some time before the inflaton decay when the
energy density of the universe was $\rho_{\mathrm{i}}$, and after
decay when the energy density is $\rho_{\mathrm{f}}$, is
\begin{equation}
\label{2.3}
 \mathcal{N} = \int^{\rho_{\mathrm{f}}}_{\rho_{\mathrm{i}}} \frac{H\, d\rho}{\dot{\rho}}
 = \frac{1}{3} \log \frac{\rho_{\mathrm{i}}}{\rho_{\mathrm{dec}}} + \frac{1}{4}
 \log \frac{\rho_{\mathrm{dec}}}{\rho_{\mathrm{f}}}
,
\end{equation}
where an overdot denotes a time derivative, and $\rho_{\mathrm{dec}}$ is
the energy density of the universe at 
the inflaton decay. Considering the inflaton to decay when $H =
\Gamma$ with $\Gamma$ being the decay rate for the inflaton, and since
$\rho_{\mathrm{i}}$ and $\rho_{\mathrm{f}}$ can be chosen
independently of $\Gamma$, (\ref{2.3}) can further be written as
\begin{equation}
 \mathcal{N} = \mathrm{const.} - \frac{1}{6} \log \Gamma,
\end{equation}
where we have denoted terms that are independent of $\Gamma$ by
$\mathrm{const.}$ Since~$\Gamma$ is a function of the modulus
field~$\sigma$, the curvature perturbations can be computed by
differentiating $\mathcal{N}$ in terms of~$\sigma_*$, where the
subscript~$*$ denotes values at the time when the CMB scale exits the
horizon during inflation.  Using (\ref{deltasigma}), one obtains the
power spectrum of the curvature perturbations as
\begin{equation}
 \mathcal{P}_\zeta  = \left(\frac{ \partial \mathcal{N}}{\partial
		       \sigma_*}  \right)^2 
 \mathcal{P}_{\delta \sigma_*}
= \left( \frac{\Gamma'}{6 \Gamma} \right)^2
 \left(  \frac{H_*}{2 \pi} \right)^2,
 \label{Pzeta}
\end{equation}
where the prime denotes a derivative with respect to $\sigma_*$.
In this mechanism, primordial non-Gaussianity can be large and the
non-linearity parameter~$f_{\mathrm{NL}}$, which characterizes the
amplitude of three point correlation function for $\zeta$, is given by
\cite{Zaldarriaga:2003my,Suyama:2007bg,Ichikawa:2008ne},
\begin{equation}
 f_{\mathrm{NL}}  = 
 \frac{5}{6} \frac{\partial^2 \mathcal{N}}{\partial
  \sigma_*^2} 
 \left( \frac{\partial \mathcal{N}}{\partial \sigma_*} \right)^{-2}
= 5 \left(1 - \frac{\Gamma\Gamma''}{\Gamma'^2}   \right).
 \label{fNL}
\end{equation}

\section{Dark Radiation from Modulated Reheating}

As we already stressed, the modulus necessarily couples to the
inflaton in order to affect the inflaton decay rate.  Further when the
modulus is light enough to be frozen at some field value until the
inflaton decay, its super-horizon field fluctuations induce the
modulated decay and produce cosmological perturbations. (If the
modulus mass was heavier than the Hubble parameter at the inflaton
decay, then the modulus would oscillate about its potential minimum
and the super-horizon fluctuations would be suppressed.) Since such
modulus should be much lighter than the inflaton which already has
been oscillating prior to its decay, and in addition, it only has
highly suppressed couplings with the SM particles because otherwise it
would receive thermal mass from them and hence would start its
oscillation before inflaton decay, the modulus serves as a
natural candidate for dark radiation.  In this section we lay out the
conditions for the modulus to be the dark radiation. Here we discuss
the case of a rather general coupling~(\ref{coupling}). Then in the
next section we present an explicit example for the
function~$f(\sigma)$.  In order to avoid isocurvature perturbations,
we assume that the inflaton decay and the production of the SM
particles/modulus to proceed only through the
interaction~(\ref{coupling}), although this assumption is not a
sufficient condition for vanishing isocurvature perturbations, as we
will soon see. All the decay products, i.e. the SM particles and the
modulus, are assumed to be relativistic at the inflaton decay.

We divide the modulus field as
\begin{equation}
 \sigma = \sigma_c + \hat{\sigma},
\end{equation}
where $\sigma_c$ is the classical background, and $\hat{\sigma}$
denotes the quantum fluctuations around it.  The interaction
term~(\ref{coupling}) can now be expanded as
\begin{equation}
 \phi f(\sigma) \mathcal{O}_{\mathrm{SM}} = \phi f(\sigma_c)
  \mathcal{O}_{\mathrm{SM}} + \phi f^{(1)}(\sigma_c) \hat{\sigma}
  \mathcal{O}_{\mathrm{SM}} + \phi f^{(2)}(\sigma_c)
  \frac{\hat{\sigma}^2}{2} \mathcal{O}_{\mathrm{SM}} + \cdots,
 \label{expand}
\end{equation}
where in the right hand side the term with an $i$-th
derivative~$f^{(i)}(\sigma_c)$ 
represents the inflaton decay channel into the SM particles and $i$
modulus particles.  Since $\sigma_c$ possesses super-horizon field
fluctuations~$\delta \sigma_c$ as in~(\ref{deltasigma}), that is,
$\sigma_c = \sigma_0+\delta\sigma_c$ ($\sigma_0$ : homogeneous
mode),\footnote{
  Note that even though the fluctuations were originally sourced as
  quantum fluctuations during inflation, they can be considered as
  classical after stretched to super-horizon sizes.
} the decay rates of all the channels fluctuate as well.  For
quantities (say,~$x = x_0 + \delta x$) discussed in this paper, we
assume that their fluctuations always satisfy $|x| \gg |\delta x|$,
so that $x_0$ can be considered as the spatially averaged value in the universe.
This is in many cases expected to hold as long as the required tuning
of $\sigma_c$ to specific values are no less than~$\delta \sigma_c$.

Referring to the decay channel which produces $i$ modulus particles as
the $i$-channel, its decay rate follows
\begin{equation}
 \Gamma_i \propto \epsilon^i \left[ (m_\phi)^i  f^{(i)}(\sigma_c)  \right]^2,
 \label{Gamma_i}
\end{equation}
where $\epsilon \sim 10^{-3}$ represents the phase space factor and 
$m_\phi$ is the mass of the oscillating inflaton.
The $i$-th derivative $f^{(i)}(\sigma_c) $ is compensated by powers
of~$m_{\phi}$, thus fixing the overall mass dimension of the decay rate
(only the $m_{\phi}$-dependence relevant to $i$ is given in the
expression). Hence the fluctuation 
of the decay rate at the linear order in~$\delta \sigma_c$ is
\begin{equation}
 \delta \Gamma_i \propto 2 \epsilon^i (m_\phi)^{2i} f^{(i)}(\sigma_0)
  f^{(i+1)}(\sigma_0) \delta \sigma_c.
 \label{delGamma_i}
\end{equation}
Here one clearly sees that the decay rate of each channel can
fluctuate in different manners, thus sourcing adiabatic as well as
isocurvature perturbations.  See Appendix~\ref{app:iso} for detailed
discussions on this issue.  In order to suppress the isocurvature
perturbations, we consider the case where one of the channels (say,
the $i$-channel) is dominant over others, i.e. has the largest decay
rate. (This corresponds to Solution~\ref{item2} at the end of the
appendix.)  In the following subsection we will discuss how dominant
the $i$-channel should be in order to satisfy observational
constraints.

\vspace{\baselineskip}

Before moving on, let us briefly comment on another way to avoid
isocurvature perturbations: When all the decay channels are modulated
in a similar fashion, i.e. $\delta \Gamma_i / \Gamma_i$ being
independent of~$i$, then isocurvature perturbations vanish at the
linear order.  (This is discussed as Solution~\ref{item1} in the
appendix.) This can be realized by a function~$f(\sigma)$ satisfying
\begin{equation}
f^{(1)}/f = f^{(2)}/f^{(1)} = f^{(3)}/f^{(2)} = \cdots \label{fff},
\end{equation}
at $\sigma = \sigma_0$.  For example, a function with the following
form satisfies the requirement:
\begin{equation}
 f(\sigma) = \exp\left( \frac{\sigma}{M} \right)
\end{equation}
with $M$ being a constant.  However, in order to generate dark
radiation corresponding to \mbox{$\Delta N_{\mathrm{eff}} \sim 1$},
the $0$-channel producing only the SM particles and the other channels
should distribute the inflaton energy to the SM particles and the
modulus at the same order.  In such case the decay rates of all the
channels are expected to be of the same order, which invalidates the
expansion~(\ref{expand}). Since channels with sufficiently large~$i$
produce more modulus particles than the SM particles, at the end the
production of the modulus may dominate over that of the SM particles.
It would be interesting to further investigate possibilities of a
model with ~(\ref{fff}), however in the following sections we focus on
the case where there is effectively only one decay channel.

\subsection{Adiabatic and Isocurvature Perturbations}
\label{subsec:AIP}

As we have discussed above, we consider the inflaton to decay into
radiation mainly through the $i$-channel, whose decay rate~$\Gamma_i$
is proportional to the square of $f^{(i)}(\sigma_0)$
(see~(\ref{Gamma_i})).  We suppose that the energy density of the
classical background~$\sigma_c$ is negligibly tiny compared to the
total energy density of the universe.  Since the modulus is assumed to
be sufficiently light so that it is frozen to a certain field value at
least until the inflaton decay, we can set the modulus field value to
be the same at the inflaton decay and at the CMB scale horizon exit,
i.e. $\sigma_0 = \sigma_*$, and also $\delta \sigma_c = \delta
\sigma_* = H_*/ 2 \pi$.  Hence the power spectrum of the adiabatic
perturbations~(\ref{Pzeta}) and its non-linearity
parameter~(\ref{fNL}) are obtained as
\begin{align}
 \mathcal{P}_{\zeta} & = \left(\frac{1}{3} \frac{f^{(i+1)}
			(\sigma_0)}{f^{(i)} (\sigma_0)}\right)^2 
 \left( \frac{H_*}{2 \pi }\right)^2
 \approx 2.4 \times 10^{-9},
 \label{Pzetaf}
\\
 f_{\mathrm{NL}} & = \frac{5}{2} 
 \left( 1 - \frac{f^{(i)} (\sigma_0) f^{(i+2)} (\sigma_0)}{f^{(i+1)}
  (\sigma_0)^2} \right).
\label{fNLf}
\end{align}
In~(\ref{Pzetaf}), the WMAP normalization value~\cite{Komatsu:2010fb}
is also shown.

\vspace{\baselineskip}

In fact, when the modulus produced from the inflaton decay 
  plays the role of dark radiation, isocurvature fluctuations in the dark
  radiation sector could possibly be generated, whose size is
  characterized by 
\begin{equation}
\label{eq:S_DR}
S_{\rm DR}  = 3 (\zeta_{\rm DR}  - \zeta_r ).
\end{equation}
Here $\zeta_\alpha$ is the curvature perturbation on the uniform energy
density hypersurface of species~$\alpha$.  Too large isocurvature
fluctuations would be inconsistent with the CMB observations and thus
severely constrained \cite{Kawasaki:2011rc}.  By using the data from
CMB, BAO and the direct measurement of the Hubble constant, for the
case where  adiabatic and isocurvature perturbations are totally
(anti-)correlated, which is relevant to our case~\footnote{
\label{ft}
In this paper we consider $\delta \sigma$ to be the only source of
density perturbations, and other contributions (e.g. from the inflaton)
to be negligibly tiny. 
Hence any isocurvature perturbations are correlated with the curvature
perturbations. 
The correlation can be positive or negative, depending on the function
$f(\sigma)$.}, the 2$\sigma$ limit on 
the power spectrum for $\zeta_{\rm DR}$, denoted as
$\mathcal{P}_{\mathcal{S}\, \mathrm{DR}} $, is roughly given by
\cite{Kawasaki:2011rc}
\begin{equation}
 \frac{\mathcal{P}_{\mathcal{S}\, \mathrm{DR}}}{\mathcal{P}_\zeta } \lesssim
  0.01,
\label{obscon}
\end{equation}
when  $\Delta N_{\mathrm{eff}} \sim 1$.

We assume that the ratio between the energy densities for the SM
particles and modulus produced through the dominant $i$-channel is of
order unity (so that the weights introduced in Appendix~\ref{app:iso}
satisfy $c_{\mathrm{SM} i}/c_{\sigma i} = O(1)$), and also that such
ratios in all channels are constants without any spatial
fluctuation. Then based on the discussions in Appendix~\ref{app:iso}
(and especially from~(\ref{A.5})), the isocurvature constraints are
satisfied if the decay rates and their fluctuations for the
subdominant channels $j$ ($\neq i$) are suppressed as
\begin{equation}
 \left|\frac{\Gamma_j }{\Gamma_i}  \right| , \, 
 \left| \frac{ \delta \Gamma_j}{\delta \Gamma_i}  \right|
\lesssim 0.1.
 \label{0.1}
\end{equation}
Hence from (\ref{Gamma_i}) and (\ref{delGamma_i}), the isocurvature
constraint~(\ref{obscon}) is satisfied under
\begin{equation}
\epsilon^{j-i}
 \left(
 \frac{f^{(j)} (\sigma_0)}{f^{(i)} (\sigma_0)}m_\phi^{j-i}
 \right)^2, \,
 \epsilon^{j-i}
 \left|
 \frac{f^{(j)} (\sigma_0)  f^{(j+1)} (\sigma_0) }{f^{(i)} (\sigma_0)
 f^{(i+1)} (\sigma_0) } m_\phi^{2(j-i)}
 \right|
 \lesssim 0.1,
 \label{2}
\end{equation}
for every $j$ with$j \neq i$. 

\subsection{Modulus as Dark Radiation}

The abundance of $\sigma$ particles (dark radiation) in terms of the
extra effective number of neutrinos is given by
\begin{equation}
 \frac{\rho_{\hat{\sigma}}}{\rho_\nu } = \Delta N_{\mathrm{eff}},
\end{equation}
where $\rho_{\hat{\sigma}}$ and $\rho_\nu$ denote the energy densities
of $\sigma$ particles (dark radiation) and one generation of neutrino,
respectively . Note that, after the neutrino decoupling, both the
energy densities of dark radiation and neutrino decrease in the same
way, i.e. proportional to $a^{-4}$, hence the extra effective number $\Delta
N_{\rm eff}$ is fixed. At the neutrino decoupling (at
temperature~$T=T_D$), the ratio of 
the energy densities of one-generation neutrino and the relativistic
SM particles is estimated as
\begin{equation}
   \frac{\rho_{r}(T_D)}{\rho_\nu(T_D)} = \frac{43}{7}.
\end{equation}
On the other hand, the energy densities of the relativistic SM
particles at the neutrino decoupling ($T=T_D$) and reheating ($T=T_R$) are
related as
\begin{equation}
   \frac{\rho_{r}(T_R)}{\rho_r(T_D)} = 
    \left( \frac{g_{*}(T_R)}{g_{*}(T_D)} \right)^{-\frac13}
    \left( \frac{a(T_R)}{a(T_D)} \right)^{-4},
\end{equation}
where $g_{*}(T)$ is the relativistic degree of freedom of the SM
particles and $a(T)$ is the scale factor at $T$.
 
Setting the number of the SM particles produced in the dominant
$i$-channel to be~$k$ (being the same order as~$i$), the resulting
energy density ratio between the modulus and SM particles at the decay
of the inflaton is
\begin{equation}
 \left. \frac{\rho_{\hat{\sigma}}}{\rho_{r}}
 \right|_{\rm inflaton~decay}  \simeq \frac{i}{k} .
 \label{ik}
\end{equation}
Here for simplicity we have assumed that the inflaton energy is
equally distributed to each decay particle, and neglected further
complication such as the momentum distribution of the decay products. Also,
energy input through the other channels are omitted (which give uncertainty 
of order 10\% at most, cf.~(\ref{0.1})). Assuming the SM particles are quickly
thermalized after the inflaton decay, $\left. \rho_{\rm SM}
\right|_{\rm inflaton~decay} \simeq \rho_{r}(T_R)$ and
$\left. \rho_{\hat{\sigma}} \right|_{\rm inflaton~decay} \simeq
\rho_{\hat{\sigma}}(T_R)$.

Then, the extra effective number $\Delta N_{\rm eff}$ of the produced
modulus (dark radiation) after the neutrino decoupling can be
estimated as
\begin{eqnarray}
  \Delta N_{\rm eff} &=&
   \frac{\rho_{\hat{\sigma}}(T_D)}{\rho_{\nu}(T_D)} \nonumber \\
    &=& \frac{\rho_{\hat{\sigma}}(T_D)}{\rho_{\hat{\sigma}}(T_R)}
      \frac{\rho_{\hat{\sigma}}(T_R)}{\rho_{r}(T_R)}
      \frac{\rho_{r}(T_R)}{\rho_{r}(T_D)}
      \frac{\rho_{r}(T_D)}{\rho_{\nu}(T_D)} \nonumber \\
    &=& \left( \frac{a(T_D)}{a(T_R)} \right)^{-4}
      \frac{i}{k}
      \left( \frac{g_{*}(T_R)}{g_{*}(T_D)} \right)^{-\frac13}
      \left( \frac{a(T_R)}{a(T_D)} \right)^{-4}
      \frac{43}{7} \nonumber \\
    &=& \frac{43}{7} \frac{i}{k} \left( \frac{g_{*}(T_R)}{g_{*}(T_D)}
			       \right)^{-\frac13}. \label{3}
\end{eqnarray}
An explicit number will be given when we discuss a
concrete model in the next section. 

\vspace{\baselineskip}

We also lay out the condition for the modulus to serve as 
dark radiation at least until the last
scattering.\footnote{
If the dark radiation (which corresponds to the extra
  neutrino numbers~$\Delta N_{\mathrm{eff}}$) turns into (a small fraction of) dark matter
  at redshift~$z_c$ after last scattering (at $z_{\mathrm{ls}}$), then
  its present abundance is written in terms of that of the photons as
\begin{equation}
 \Omega_{\mathrm{DR}\to \mathrm{DM}\, 0} \simeq 
 0.2 \times \Delta N_{\mathrm{eff}}\,  z_c\,  \Omega_{\gamma0} 
 < 0.2 \times \Delta N_{\mathrm{eff}}\,  z_{\mathrm{ls}}\,  \Omega_{\gamma0} 
 \sim
 0.01 \times \Delta N_{\mathrm{eff}}.
\end{equation}
Hence as long as $\Delta N_{\mathrm{eff}} = \mathcal{O}(1)$, their
present abundance is tiny compared to  the total dark matter abundance,
$\Omega_{\mathrm{DM}0} \approx 0.22$.} The modulus particle
produced through the 
$i$-channel obtains momentum of $m_\phi / (i+k)$,
thus in order for this particle to stay
relativistic until last scattering, the modulus
mass~$m_\sigma$ is required to be as small as
\begin{equation}
 \frac{m_\phi}{i+k} \frac{a_{\mathrm{dec}}}{a_{\mathrm{ls}}}
 \sim \frac{m_{\phi}}{3 (i+k)}
  \left(\frac{H_{\mathrm{eq}}}{\Gamma_i}\right)^{1/2} \gg m_{\sigma} ,
\label{D4}
\end{equation}
where $a_{\mathrm{dec}}$ denotes the scale factor at the inflaton
decay, $a_{\mathrm{ls}}$ the scale factor at last scattering, and 
$H_{\mathrm{eq}}$ the Hubble parameter at matter-radiation equality.

\subsection{Oscillating Modulus as Dark Matter}
\label{subsec:OMDM}

So far we have not cared about the fate of the classical
background~$\sigma_c$.  In this subsection, we further investigate the
possibility that the classical background of the modulus starts 
sinusoidal oscillations after the inflaton decay and becomes the dark
matter. Note that still in such case, the quantum
fluctuation~$\hat{\sigma}$ can serve as the dark radiation, given that
it obtains enough momentum at the inflaton decay.
This sets an lower bound on the inflaton mass (cf.~(\ref{D4})), 
but as we will see shortly, the bound
is satisfied for almost all the models since the mass of $\sigma$
required to explain the dark matter abundance 
is extremely small.\footnote{Such an extremely light modulus may be ubiquitous in the Axiverse scenario~\cite{Arvanitaki:2009fg}.}

For simplicity, we consider the modulus potential to be of the
quadratic type
\begin{equation}
V(\sigma) = \frac{1}{2} m_\sigma^2 (\sigma -
\sigma_{\mathrm{min}})^2, 
\end{equation}
where $m_\sigma$ is the modulus mass and $\sigma_{\mathrm{min}}$ is
the potential minimum. Hence the classical background and quantum
fluctuation~$\hat{\sigma} $ both have the same mass~$m_\sigma$, and
the background starts its sinusoidal oscillations from the field
value~$\sigma_c$ when $H \sim m_{\sigma}$. Here we introduce
\begin{equation}
 \sigma_{\mathrm{osc}} \equiv | \sigma_c - \sigma_{\mathrm{min}}|,
\end{equation}
which represents the initial oscillation amplitude of the modulus
background. In the following we give rough estimations
for the modulus to serve both as dark radiation as well as dark
matter.

\subsubsection*{Dark Matter Abundance}

The present abundance of dark matter from the oscillating
modulus, whose value should be $ \Omega_{\mathrm{DM}0} \approx 0.22$
from current observations, is estimated as
\begin{equation}
 \Omega_{\mathrm{DM} 0} 
= \frac{1}{\rho_c}
 \frac{s_0}{s_{\mathrm{osc}}}
 \frac{1}{2 }  m_\sigma^2 \sigma_{\mathrm{osc}}^2
 \simeq
 0.1 \left(\frac{m_\sigma}{10^{-27} \mathrm{eV}}  \right)^{1/2}
 \left( \frac{\sigma_{\mathrm{osc}}}{M_p}  \right)^2,
\end{equation}
where $M_p$ denotes the reduced Planck mass, 
$\rho_c$ is the present critical density, $s_0$ and $s_{\mathrm{osc}}$
are the entropy densities of the SM particles at present and when the
oscillation starts. Upon estimating~$s_{\mathrm{osc}}$ and obtaining the
far right hand side, we have assumed the onset of the oscillation to be
at around or after the neutrino decoupling, and that $\Delta
N_{\mathrm{eff}} \lesssim 1$. (However we note that the result is not
sensitive to such assumptions, e.g. if $g_* \sim 200$
at the onset of the oscillation, then the numerical factor in the
far right hand side is about $ 0.06$ for $\Delta N_{\mathrm{eff}} =
1$.)

Furthermore, we require dark matter to have been present at least by
the time of matter-radiation equality.  This sets the latest time the
modulus background starts its oscillation, giving a lower bound on the
modulus mass,
\begin{equation}
 m_\sigma > H_{\mathrm{eq}} \sim 2 \times 10^{-28} \,{\rm eV}.
 \label{D2}
\end{equation}

On the other hand, the upper bound on the modulus mass was given
in~(\ref{D4}) from the requirement that the quantum
fluctuation~$\hat{\sigma}$ produced by the inflaton decay to serve as
dark radiation at least until the last scattering.
The condition (\ref{D4}) can be rewritten in terms of $T_R$ and $m_\sigma$ as
\beq
m_\phi\;\gg\; \GEV{-17}\, (i+k) 
\left(\frac{g_*(T_R)}{228.75}  \right)^{1/4}
\lrf{T_R}{\GEV{9}} \lrf{m_\sigma}{10^{-27}\,{\rm eV}}.
\label{constraint}
\eeq
Here, $g_* = 228.75$ corresponds to the maximum value allowed in the
MSSM, and we have used the approximation, 
\beq
\Gamma_i \simeq \lrfp{\pi^2 g_*(T_R)}{90}{\frac{1}{2}} \frac{T_R^2}{M_p}.
\eeq

Therefore there is a wide range of the parameters where (\ref{D2}) and (\ref{constraint}) are
satisfied.

\vspace{\baselineskip}

We note that the potential energy of the modulus is negligibly small
compared to the total energy of the universe upon the inflaton decay,
as long as the oscillation amplitude is smaller than the Planck scale.
This validates our computations of the perturbations in the previous
sections where we have neglected the energy density of~$\sigma_c$.
Note that this is not always the case especially if one considers
a potential deviated from the simple quadratic potential.

%

\subsubsection*{Dark Matter Isocurvature}

In addition to the adiabatic perturbations~(\ref{Pzetaf}), the
oscillating modulus obtains isocurvature perturbations of order
\begin{equation}
 \frac{\delta \rho_{\mathrm{DM}}}{\rho_{\mathrm{DM}}} \simeq
 \frac{\delta \sigma_c}{ \sigma_{\mathrm{osc}}},
\label{DMisosigma}
\end{equation}
which is necessarily correlated with the curvature perturbation (see footnote \ref{ft}).
The correlation can be positive or negative, depending on the position of the potential minimum.
To be concrete, let us adopt the observational constraint on 
the totally anti-correlated dark matter isocurvature perturbations~\cite{Komatsu:2010fb},
roughly given by
\begin{equation}
  \frac{\mathcal{P}_{\mathcal{S}\, \mathrm{DM}}}{\mathcal{P}_\zeta } \lesssim
  0.01,
\end{equation}
where $\mathcal{P}_{\mathcal{S}\, \mathrm{DM}}$ is the power spectrum 
of dark matter isocurvature fluctuations $S_{\rm DM}$ defined in the same manner as in \eqref{eq:S_DR} 
by replacing $\zeta_{\rm DR}$ with $\zeta_{\rm DM}$.

Hence we require the initial oscillation amplitude of the modulus as
\begin{equation}
 \frac{1}{\sigma_{\mathrm{osc}}} \frac{f^{(i)} (\sigma_0)}{f^{(i+1)}
  (\sigma_0)} \lesssim 0.01
 \label{D1}
\end{equation}
in order to suppress the dark matter isocurvature perturbations. 

Here we note that the dark matter isocurvature
perturbation~(\ref{DMisosigma}) can be absent if the modulus possesses
a non-quadratic potential. As was indicated in~\cite{Kawasaki:2011pd}
for the case of the curvaton
mechanism~\cite{Linde:1996gt,Enqvist:2001zp,Lyth:2001nq,Moroi:2001ct},
a non-quadratic potential sources non-uniform onset of the field
oscillation, which can cancel out the
perturbation~(\ref{DMisosigma}). 
We also remark that for such
non-quadratic potentials the classical background and the quantum
fluctuation around it can possess different masses.
Moreover, the initial mass of the classical background (which sets the
starting time of the oscillation) can be quite different from the final
mass for the oscillation.

\section{A Concrete Example}

Let us illustrate in a toy example how the modulus serves as dark
radiation with the abundance $\Delta N_{\mathrm{eff}} \sim 1$.
The coupling term studied in this section is of the form
\begin{equation}
 \mathcal{L} \supset \kappa \frac{ \phi}{M_p} \frac{\sigma}{M}
  e^{\sigma/M} F_{\mu\nu} F^{\mu\nu},
\end{equation}
where $\kappa$ is a dimensionless coupling constant, $M$ a constant of
mass dimension, and $F_{\mu\nu} = \partial_\mu A_\nu - \partial_\nu
A_\mu$ denotes the gauge field strength of $U(1)_{\rm EM}$.  The
inflaton~$\phi$ decays into $i$ modulus particles and $2$ photons ($k
= 2$), where the tree-level decay rates of the channels are
\begin{equation}
\begin{split}
 \Gamma_0 (\phi \to 2 \gamma) & = 
\frac{1}{4 \pi }
 \left( \frac{\kappa }{M_p} \frac{\sigma_c}{M} e^{\sigma_c / M}
 \right)^2 m_\phi^3,  \\
 \Gamma_1 (\phi \to \hat{\sigma} + 2 \gamma) & = 
 \frac{1}{768 \pi^3} 
 \left\{\frac{\kappa }{M_p} \left(1 + \frac{ \sigma_c}{M}\right)
 e^{\sigma_c / M}
 \right\}^2 \frac{m_\phi^5 }{M^2}, 
 \\
 & \, \, \, \vdots
\end{split}
\end{equation}
When $|\sigma_0|$ is sufficiently smaller than $m_\phi$ and $M$, and 
$m_\phi$ not so large compared to~$M$, then
the $i=1$ channel becomes dominant
over the others.  (More generally, the $n$-channel can be made
dominant in a similar fashion by couplings of the form $\phi \sigma^n
e^{\sigma/M} \mathcal{O}_{\mathrm{SM}}$.)  Then the adiabatic
perturbation spectrum~(\ref{Pzetaf}) is of the form
\begin{equation}
 \mathcal{P}_\zeta \simeq \left( \frac{1}{3 \pi} \frac{H_*}{M}
			  \right)^2 \approx 
 2.4 \times 10^{-9},
\end{equation}
with order-unity non-linearity parameter~(\ref{fNLf}),
\begin{equation}
 f_{\mathrm{NL}} \simeq \frac{5}{8}.
\end{equation}
Moreover, the isocurvature constraints~(\ref{0.1}) (or~(\ref{2})) are
satisfied under
\begin{equation}
 \frac{|\sigma_0|}{m_\phi}\lesssim 10^{-2} , \qquad
\frac{|\sigma_0|}{M} \lesssim 10^{-4}  \frac{m_\phi^2}{M^2} , \qquad
 \frac{m_\phi^2}{M^2} \lesssim
 10 . 
\end{equation}
Here we also note that $|\sigma_0|$ is considered to be no less than
the super-horizon fluctuations~$H_{\mathrm{inf}} / 2 \pi$.

The abundance of the dark radiation~$\hat{\sigma}$ in terms of the
effective number of neutrinos~(\ref{3}) is obtained as
\begin{equation}
 \Delta N_{\mathrm{eff}} \simeq 
    \frac{43}{7}\cdot \frac{1}{2} \cdot\left( \frac{g_{*}(T_R)}{g_{*}(T_D)}
			     \right)^{-\frac13},
\end{equation}
which yields $\Delta N_{\rm eff} \sim 1.1$ for $g_{\ast}(T_D) \simeq
10.75$ and $g_{\ast}(T_R) \simeq 228.75$.

\vspace{\baselineskip}

Conditions under which the oscillating classical background of the
modulus serves as dark matter can be obtained straightforwardly
following the discussions in Subsection~\ref{subsec:OMDM}. Let us just
point out here that the isocurvature constraint~(\ref{D1}) requires a
rather large oscillation amplitude for the modulus such that
$\sigma_{\mathrm{osc}} > 10 M$.

\section{Discussion and Conclusions}
\label{sec:3}

In this paper, we have discussed cosmological implications of the
  modulus in the modulated reheating mechanism, which is one of the alternative mechanism generating 
  primordial fluctuation, and studied especially in connection with primordial
  non-Gaussianity.  Since the modulus has to couple to the inflaton in
  order to modulate the inflaton decay rate, it is inevitably
  produced by the decay.  It also should be rather light and have only
  very suppressed interactions with the SM particles for successful
  modulated mechanism, thus, the modulus is a natural candidate
  for dark radiation. The presence of the dark radiation has a lot of
  implications on BBN and CMB. It speeds up the expansion of the
  Universe and hence the Helium abundance is increased. As for the CMB
  power spectrum, there are mainly three effects
  \cite{Bashinsky:2003tk,Ichikawa:2008pz}: (A) The matter-radiation
  equality is delayed so that the early ISW effect is enhanced, which
  leads to the increase of the height of the first acoustic peak, then
  relative suppression of other higher peaks. (B) The free-streaming
  effects become more significant so that the amplitudes at small
  angle scales are suppressed. (C) The sound horizon becomes smaller due
  to the additional contribution to the Hubble parameter at the
  recombination, which shifts the peaks and troughs to smaller scales. 
  Recent observations might have seen these kinds of effects,
  in particular on small scales, which suggest
  $\Delta N_{\rm eff}\sim 1$.  It is also interesting to note that when
  $\Delta N_{\rm eff} > 0$, more scale-invariant or slightly bluer
  spectral index $n_s$ is favored compared to the case with $\Delta
  N_{\rm eff} =0$ where slightly red-tilted spectrum is favored.\footnote{
    The analysis with the data from WMAP+SPT shows $n_s = 0.9663
      \pm 0.0112$ (1$\sigma$) for $\Lambda$CDM model with a fixed
      $N_{\rm eff} = 3.046$, on the other hand, when $N_{\rm eff}$ is
      varied, the constraint becomes $n_s = 0.9874 \pm
      0.0193$~(1$\sigma$)~\cite{Keisler:2011aw}. Thus the
      scale-invariant spectrum is consistent with observation within $1
      \sigma$, in the presence of the extra radiation.
    It is worth noticing that the modulated reheating mechanism predicts $n_s -1 = -
  2\epsilon + 2 \eta_\sigma$ where $\epsilon = -\dot{H}/H^2$ and
  $\eta_\sigma = V^{\prime\prime}/ 3 H^2$,
  cf.~(\ref{spectralindex}).}
Since the modulus mass is required to be extremely light for 
modulated reheating to operate, the resulting perturbation spectrum
tends to be very flat (except for cases with specific inflationary
mechanisms, see also discussions in footnote~\ref{foot123} and
Appendix~\ref{app:SI}). 
The modulated reheating mechanism shows a good agreement with the
current data in this respect as well, in addition to giving a explicit
candidate for dark radiation.

In general, produced modulus particles (dark radiation) have
isocurvature fluctuations, which are severely constrained from
observations. We have derived general conditions to suppress
isocurvature perturbations in the modulated reheating mechanism. Such
conditions are easily satisfied in the following three cases: (i) The
ratios of the decay rates to their fluctuations for all decay channels
are almost the same. (ii) One decay channel dominates over the
others. (iii) The weights $c_{\sigma_i}, c_{\chi_i}, \cdots$ have
special relations between them. As a concrete example, a toy model
corresponding to the case (ii) has been discussed. In the model, the
decay mode with one modulus particle indeed dominates over the other
modes and the dark radiation with $\Delta N_{\rm eff} \sim 1$ is
realized.
Though we have mainly focused on the case (ii) in this
paper, it would be interesting to investigate other cases in detail.

While modulus particles generated by inflaton decay serve as dark
radiation, coherent oscillations of modulus can behave like dark
matter as well if its mass is adequately large and it can start
oscillation before the matter-radiation equality. We have shown that
both the dark radiation and the CDM can be attributed to the modulus
by taking appropriate parameters.

Lastly we comment on a case where the modulus has a heavier mass, and
does not account for dark radiation. 
Such modulus produced by the inflaton decay
may also have an impact on cosmology. For the case of a stable modulus,
its abundance should not exceed the observed dark matter abundance.
If it is unstable and decays into the SM particles, the energetic decay
products may significantly change  the light element abundance in contradiction
with observation. Thus, even if the modulus
does not account for dark radiation, it will be worth studying the cosmological
effects of the modulus, which is necessarily produced by the inflaton decay
in the modulated reheating mechanism.

\section*{Acknowledgements}

TK would like to thank Amir Hajian for helpful conversations. 
FT thanks W. Buchm\"{u}ller and the DESY theory group for the warm
hospitality while the present work was completed.
  This work was supported in part by JSPS Grant-in-Aid for Scientific
Research Nos.~21740187 (M.Y.), 23740195 (T.T.).
This work was also supported by the Grant-in-Aid for Scientific Research on Innovative
Areas (No.21111006) [FT], Scientific Research (A) (No.22244030 [FT] and No.21244033 [FT]), and JSPS Grant-in-Aid
for Young Scientists (B) (No.21740160) [FT].  
The work of TT is partially supported by Saga University Dean's Grant 2011 For Promising
Young Researchers. 
This work was also supported by World Premier International Center Initiative (WPI
Program), MEXT, Japan.

\appendix

\section{Isocurvature Perturbations in Modulated Reheating}
\label{app:iso}

In this appendix we give general discussions on isocurvature
perturbations among particles created through modulated decay.  We
consider the universe to be initially dominated by a single
matter-like component~$\phi$ (e.g. the inflaton), which decays into
multiple relativistic components~$\sigma$, $\chi$, $\ldots$, through
multiple decay channels.  Each decay particle is produced by one or
more of the decay channels, and the decay rates of (some of) the
channels possess super-horizon fluctuations.  Interactions between the
decay components are assumed to be negligibly tiny, hence
thermalization among the components are forbidden.

Now let us focus on one of the decay particles, say,~$\sigma$, and
estimate its energy density fluctuations~$\delta \rho_\sigma$ due to
the modulated decay. We represent the decay rate of channel~$i$ by
$\Gamma_i$, and its fluctuations (if any) by $\delta \Gamma_i$. The
size of the fluctuations $|\delta \Gamma_i|$ are assumed to be much
smaller than the values of~$|\Gamma_i|$ themselves.  Furthermore, we
denote the fractional energy density of~$\sigma$ produced through
channel~$i$ as $\rho_{\sigma i}$ (hence $\rho_\sigma = \sum_i
\rho_{\sigma i}$), and assume that the decay of $\phi$ suddenly
happens when $H = \sum_i \Gamma_i \equiv H_{\rm dec}$. Then
$\rho_{\sigma i}$ at the time when $\phi$ decays can be written in
terms of the energy density of $\phi$ as
\begin{equation}
 \rho_{\sigma i \mathrm{dec}} = \frac{c_{\sigma i}
  \Gamma_i}{\sum_j \Gamma_j} \rho_{\phi \mathrm{dec}},
\end{equation}
where the subscript ``$\mathrm{dec}$'' denotes values at the $\phi$
decay. The weight $c_{\sigma i}$ ($ 0 \leq c_{\sigma i} \leq 1$)
represents the fraction of $\rho_\phi$ that goes into $\rho_\sigma$
through channel $i$. For simplicity, we assume that $c_{\sigma i}$'s
are constants, which do not fluctuate.

At some time after the $\phi$ decay when the size of the universe
is~$a$, the energy density $\rho_{\sigma i}$ becomes
\begin{equation}
 \rho_{\sigma i} = \rho_{\sigma i \mathrm{dec}} 
 \left( \frac{a_{\mathrm{dec}}}{a}\right)^4 
 = \rho_{\sigma i \mathrm{dec}} 
 \left(\frac{a_0}{a}\right)^4 \left(\frac{H_0}{H_{\mathrm{dec}}}\right)^{8/3}.
\end{equation}
In the far right hand side we have introduced values~$a_0$ and $H_0$
which are values at some time before the decay. Let us now consider
energy density fluctuations on the spacially flat slicing, 
where the scale factors are constants.
Note that $H_0$ is also homogeneous since it is prior to the modulated
decay. Then the spatial density fluctuations are sourced by $\delta
\Gamma$ through 
\begin{equation}
 \rho_{\sigma i } \propto \Gamma_i \left(\sum_j
				    \Gamma_j\right)^{-5/3},
\end{equation}
giving, up to linear order in $\delta \Gamma_l$,
\begin{equation}
 \delta \rho_{\sigma i} \simeq \rho_{\sigma i } 
 \left(
 \frac{\delta \Gamma_i}{\Gamma_i} - \frac{5}{3} \frac{\sum_j \delta
 \Gamma_j}{\sum_k \Gamma_k }  \right).
 \end{equation}
Hence up to linear order in the decay rate fluctuations, we arrive at
\begin{equation}
 \frac{\delta \rho_\sigma}{\rho_\sigma} \simeq
 \frac{\sum_i c_{\sigma i}\,  \delta \Gamma_i}{\sum_j c_{\sigma j }\, 
 \Gamma_j} 
 - \frac{5}{3} \frac{\sum_i \delta \Gamma_i}{\sum_j \Gamma_j}.
 \label{A.5}
\end{equation}
For example, when there is only one decay channel producing
solely~$\sigma$, then $\frac{\delta \rho_\sigma}{ \rho_\sigma} \simeq
-\frac{2}{3} \frac{\delta \Gamma}{\Gamma}$.  Also, note that even if
the decay rates of all the channels producing~$\sigma$ do not have
fluctuations, i.e. $\delta \Gamma_i = 0 $ for all~$i$ satisfying
$c_{\sigma i } \neq 0$, still the energy density fluctuation~$\delta
\rho_\sigma$ is produced as long as (one of) the other decay channels
are modulated.

In order for the linear-order isocurvature fluctuations to vanish
among different decay products (which generally have different values
of~$c_i$'s), i.e. $\delta \rho_\sigma / \rho_\sigma \simeq \delta
\rho_\chi / \rho_\chi \simeq \cdots$, the first term in the right hand
side of~(\ref{A.5}) needs to take the same value for all decay
particles.  We end this appendix by laying out three possible
solutions:
\begin{enumerate}
\item Given that $\delta \Gamma_i / \Gamma_i$ take the same value for
  all decay channels, all particle products obtain the same energy
  density fluctuations. An example of such case is briefly discussed
  before Subsection~\ref{subsec:AIP}. \label{item1}
\item Another simple solution is to have only one decay channel for
  $\phi$. This case is considered in the main body of this paper.\label{item2}
\item The isocurvature perturbations vanish also when the parameters
  $c_{\sigma i}$, $c_{\chi i}$, $\ldots$ take specific values such
  that the first term in the right hand side of~(\ref{A.5}) becomes
  the same for all decay
  products. For e.g., $c_{\sigma_j}/c_{\sigma_i} =
    c_{\chi_j}/c_{\chi_i}$ for all decay products ($\sigma, \chi,
    \cdots$) and channels.
 \end{enumerate}

\section{Spectral Index in Modulated Reheating}
\label{app:SI}

In this appendix we derive the spectral index of the density
perturbation spectrum obtained from modulated reheating. 
We carry out the computations without specifying the functional forms of
the modulus potential and the inflaton decay rate, thus the results
can be applied to generic modulated reheating scenarios.

\vspace{\baselineskip}

In Section~\ref{sec:2} we derived the perturbation spectrum generated
from modulated reheating:
\begin{equation}
 P_\zeta = \left( \frac{1}{6 \Gamma } \frac{\partial \Gamma }{\partial
	    \sigma_*}  \right)^2 
 \left( \frac{H_*}{2 \pi }  \right)^2.
\end{equation}
Recall that the inflaton decay rate is a function of the
modulus, i.e.~$\Gamma = \Gamma (\sigma_{\mathrm{dec}})$, and that the
subscripts~$*$ and ``$\mathrm{dec}$'' denote values at the horizon exit of
the CMB scale, and at the inflaton decay, respectively.
Since~$\sigma_{\mathrm{osc}}$ is independent of the comoving wave
number~$k$, the spectral index of the spectrum is obtained as
\begin{equation}
 n_s -1 \equiv \frac{d \ln P_\zeta }{d \ln k}
 \simeq \frac{1}{H_*} \frac{d}{dt}\ln \left( \frac{d \, \Gamma
 (\sigma_{\mathrm{dec}})
  }{d \, \sigma_{\mathrm{dec}}}
 \frac{\partial \sigma_{\mathrm{dec}}}{\partial \sigma_*}
   \right)^2
 + 2 \frac{\dot{H}_*}{H_*^2},
 \label{B2}
\end{equation}
where an overdot denotes a time derivative, and
we have assumed the Hubble parameter during inflation to be 
almost a constant.

Upon computing the first term in the right hand side, we make use of the
following approximation for the modulus dynamics: Supposing the modulus
potential~$V(\sigma)$ to be a function only of~$\sigma$ (thus does not
explicitly depend on, say, time),
and that the dominant component of the universe obeys the equation of 
state~$p = w \rho$ with constant~$w$,
then the equation of motion of the modulus
\begin{equation}
 \ddot{\sigma} + 3 H \dot{\sigma}   = -  V'
\end{equation}
is approximated by~\cite{Chiba:2009sj} 
\begin{equation}
 c H \dot{\sigma} \simeq -V' \quad \mathrm{with} \quad
 c = \frac{3 (w + 3)}{2}, \label{chsigma}
\end{equation}
given that the effective mass of the modulus is as small as
\begin{equation}
 \left| \frac{V''}{c H^2} \right| \ll 1. \label{ccond}
\end{equation}
Note that in this appendix, a prime denotes a derivative in terms of $\sigma$
(which should not be confused with the primes in the main body of this
paper denoting $\sigma_*$-derivatives). 
(\ref{ccond}) is a necessary condition for the
approximation~(\ref{chsigma}) to hold, 
and one can further show that (\ref{ccond}) actually guarantees 
(\ref{chsigma}) to be a stable attractor. See, e.g., Appendix~A
of~\cite{Kawasaki:2011pd} for detailed discussions on scalar field
dynamics in an expanding universe.

Hence the dynamics of the modulus, whose effective mass is considered to
satisfy~(\ref{ccond}) at least until the inflaton decay, is described by
(\ref{chsigma}) with $c = 3$ during inflation (which is nothing but the
slow-roll approximation), and $c = 9/2$ after inflation until the
inflaton decay. This gives
\begin{equation}
 \int^{\sigma_{\mathrm{dec}}}_{\sigma_*} \frac{d\sigma }{V'} \simeq
 - \int^{t_{\mathrm{end}}}_{t_*} \frac{dt}{3 H} 
+ \frac{4}{27} \int^{H_{\mathrm{dec}}}_{H_{\mathrm{end}}}
\frac{dH}{H^3},
 \label{dsigmaV}
\end{equation}
where the subscript ``end'' represents values at the end of inflation.
Taking into account that the modulus has little effect on the expansion
of the universe before the inflaton decay, and that $H_{\mathrm{dec}} =
\Gamma (\sigma_{\mathrm{dec}})$, 
then after differentiating both sides of (\ref{dsigmaV}) with $\sigma_*$,
one obtains
\begin{equation}
 \frac{\partial \sigma_{\mathrm{dec}}}{ \partial \sigma_*} \simeq
 \left\{
 1 - \frac{4}{27}
 \frac{V'(\sigma_{\mathrm{dec}})}{\Gamma(\sigma_{\mathrm{dec}})^3}  
 \frac{d \,  \Gamma (\sigma_{\mathrm{dec}}) }{d \, \mathrm{\sigma_{\mathrm{dec}}}}
\right\}^{-1} \frac{V'(\sigma_{\mathrm{dec}})}{V'(\sigma_*)}.
\end{equation}
Here one sees that only the $V'(\sigma_*)$ in the denominator
contributes to the spectral index in~(\ref{B2}). Combining the above
results, we arrive at
\begin{equation}
 n_s -1 \simeq
 \frac{2}{3} \frac{V''(\sigma_*)}{ H_*^2}
 + 2 \frac{\dot{H}_*}{H_*^2}.
 \label{spectralindex}
\end{equation}

Since the effective mass of the modulus in the modulated reheating is
generically required to be very light so that (\ref{ccond}) is satisfied at least
until reheating, the resulting perturbation spectrum tends to be extremely
flat, 
unless the inflationary mechanism gives a rather large $\dot{H}_*
/H_*^2$ (see also Footnote~\ref{foot123}).

\end{document}